\documentclass{mn2e}
\usepackage{epsfig}
\usepackage{epsf}

\title[Dust Formation in Eta Carinae]{Episodic Post-Shock Dust
  Formation in the Colliding Winds of Eta Carinae}

\author[N.\ Smith]{Nathan Smith\thanks{Email:
    nathans@astro.berkeley.edu} \\ Department of Astronomy, University
  of California, Berkeley, CA 94720, USA}

\begin{document}
\date{Accepted 0000, Received 0000, in original form 0000}
\pagerange{\pageref{firstpage}--\pageref{lastpage}} \pubyear{2002}
\def\arcdeg{\degr}
\maketitle
\label{firstpage}

\begin{abstract}

  Eta Carinae shows broad peaks in near-infrared (IR) $JHKL$
  photometry, roughly correlated with times of periastron passage in
  the eccentric binary system.  After correcting for secular changes
  attributed to reduced extinction from the thinning Homunculus
  Nebula, these peaks have IR spectral energy distributions (SEDs)
  consistent with emission from hot dust at 1400--1700~K.  The excess
  SEDs are clearly inconsistent, however, with the excess being
  entirely due to free-free wind or photospheric emission.  One must
  conclude, therefore, that the broad near-IR peaks associated with
  Eta Carinae's 5.5 yr variability are due to thermal emission from
  hot dust.  I propose that this transient hot dust results from
  episodic formation of grains within compressed post-shock zones of
  the colliding winds, analogous to the episodic dust formation in
  Wolf-Rayet binary systems like WR~140 or the post-shock dust
  formation seen in some supernovae like SN~2006jc.  This dust
  formation in Eta Carinae seems to occur preferentially near and
  after periastron passage; near-IR excess emission then fades as the
  new dust disperses and cools.  With the high grain temperatures and
  Eta Car's C-poor abundances, the grains are probably composed of
  corundum or similar species that condense at high temperatures,
  rather than silicates or graphite.  Episodic dust formation in Eta
  Car's colliding winds significantly impacts our understanding of the
  system, and several observable consequences are discussed.

\end{abstract}


\section{INTRODUCTION}

As one of the brightest mid-infrared (IR) sources in the sky, Eta
Carinae has been a key reservoir of information about dust around
massive stars (Chesneau et al.\ 2005; Smith et al.\ 1998, 2003b;
Aitkin et al.\ 1995; Robinson et al.\ 1987; Hackwell et al.\ 1986).
Dust formation by massive stars is of considerable interest regarding
the presence of dust in star-forming galaxies at high redshift (e.g.,
Bertoldi et al.\ 2003), when the Universe was too young for asymptotic
giant branch stars to have contributed significantly (Tielens
1998). The neutral and dusty bipolar Homunculus Nebula, ejected in the
19th century Great Eruption, absorbs most of the ultraviolet and
visual-wavelength radiation from the star and re-radiates that
luminosity in the mid-IR, providing a handy calorimeter of the central
star (Smith et al.\ 2003b).  This paper, however, focusses attention
not on the dusty Homunculus, but on the near-IR emission from the very
core of the nebula surrounding the central star (Rigaut \& Gehring
1995; Smith \& Gehrz 2000; Chesneau et al.\ 2005), which exhibits
distinct and peculiar variability in the near-IR.

Whitelock et al. (1983, 1994, 2004) have followed the photometric
near-IR variability of Eta Carinae, and noted quasi-periodic peaks
every $\sim$5 yr.  Damineli (1996) proposed that this IR variability
and Eta Car's optical spectroscopic changes were due to an eccentric
5.5 yr binary system, and successfully predicted the time of the next
periastron passage in 1998.0.  Since then, an intensive
multiwavelength campaign has documented Eta Car's cyclical variability
in the X-rays (Corcoran 2005; Corcoran et al.\ 2001; Ishibashi et al.\
1999), optical (Damineli et al.\ 1998; Davidson et al.\ 2000), and
radio (Duncan \& White 2003).

Here we return to the near-IR emission peaks that first heralded the
$\sim$5 yr variability, which have so far eluded a satisfying
explanation.  We are interested in the broad $\Delta t \simeq$1--2 yr
peaks in the $JHKL$ light curves, and not the narrow eclipse-like
events in the month or so surrounding periastron (Whitelock et al.\
2004; Feast et al.\ 2001), which probably have a different origin and
will be the topic of a future paper.  In particular, $JHKL$ photometry
obtained from 1972 to 2004 by Whitelock et al.\ (1983, 1994, 2004) is
used to constrain the spectral energy distribution (SED) of the {\it
  excess} emission that rises above the more steady, secular increase
in apparent brightness over recent decades noted by Whitelock et al.\
(1994).  We show that the broad quasi-periodic peaks in the $JHKL$
light curves are caused by emission from hot dust grains, and we
propose that these grains condense within the dense post-shock gas in
the colliding-wind binary system.  SEDs of this excess emission are
not consistent with photospheric or free-free emission, which was
previously assumed to be the origin.


Section 2 reviews the observational data, introducing a correction for
the slow increase in apparent brightness caused by the expanding and
thinning Homunculus Nebula.  Section 3 presents the excess-emission
SEDs for several of the past periastron passage events, and discusses
possible emission mechanisms and the properties of the dust
responsible for the excess near-IR emission.  Finally, Section 4
proposes a scenario wherein new dust grains form in post-shock gas
around times of periastron in Eta Carinae as a result of compression
by the colliding winds.

\begin{figure}\begin{center}
\includegraphics[width=3.1in]{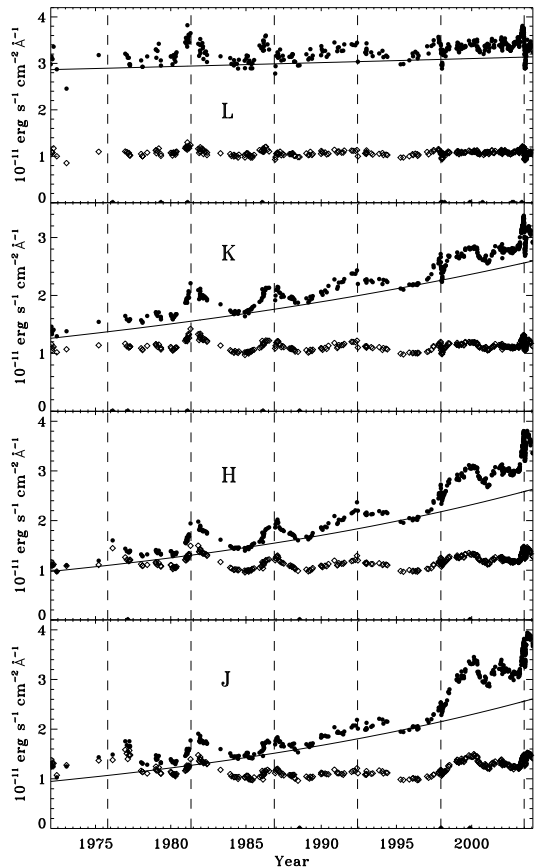}
\end{center}
\caption{Near-IR $JHKL$ photometry of Eta Car (solid points) converted
  to flux, from Whitelock et al.\ (2004).  The solid curve shows the
  adopted secular trend of the increasing continuum level, probably
  due to the expansion and thinning of the Homunculus Nebula and the
  consequent decrease in extinction to the central star.  The unfilled
  points show the photometry after dividing by this secular
  trend.}\label{fig:plotJHKL}
\end{figure}
\begin{figure}\begin{center}
\includegraphics[width=3.0in]{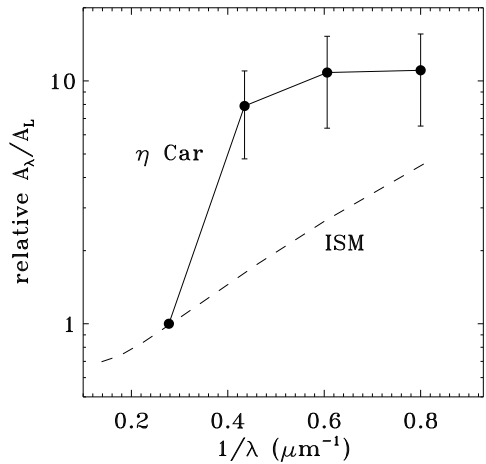}
\end{center}
\caption{Near-IR extinction law for the Homunculus of $\eta$ Car
  (solid points) normalized to the $L$-band extinction $A_L$,
  determined from the solid curves in Figure~\ref{fig:plotJHKL} that
  approximate the secular brightening trend in $JHKL$ photometry.  The
  dashed line shows the average ISM extinction law in the Milky Way
  from Indebetouw et al.\ (2005) for
  comparison.}\label{fig:extinction}
\end{figure}
\begin{figure}\begin{center}
\includegraphics[width=3.1in]{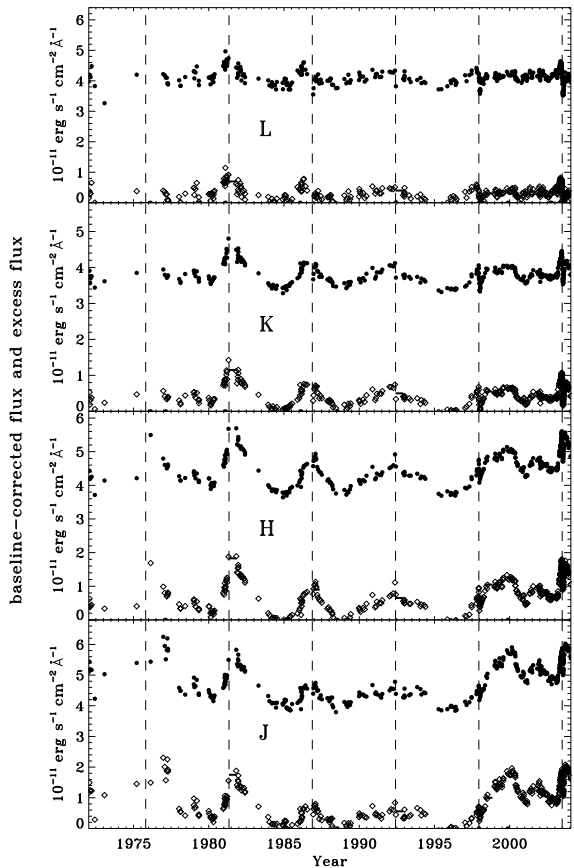}
\end{center}
\caption{Similar to Figure~\ref{fig:plotJHKL}, but the solid points
  show the near-IR flux with a correction for the slowly-changing
  extinction that is presumed to cause the secular brightening over
  this time period.  Total extinction at the latest epochs is assumed
  to be negligible.  The unfilled points show the $JHKL$ excess flux,
  with the constant baseline flux (due to stellar photospheric and
  wind emission) subtracted, leaving only the excess near-IR flux
  associated with the 5.5 yr variability. The short horizontal dashes
  show the estimated flux for each dust formation
  event.}\label{fig:plotJHKLsub}
\end{figure}

\section{NEAR-IR PHOTOMETRY}

This paper investigates near-IR ($JHKL$) photometry of Eta Carinae
collected at the South African Astronomical Observatory (SAAO) from
1972 to 2004 by Whitelock and collaborators, kindly provided to the
author by P.A.\ Whitelock.  The data represent photometry for the
entire Homunculus Nebula measured using a 36\arcsec-diameter aperture.
For details regarding the observations and calibration, see Whitelock
et al.\ (1983, 1994, 2004).  According to Whitelock et al., typical
photometric uncertainties are of order $\pm$0.03 mag at $JHK$ and
$\pm$0.05 mag at $L$. The data are invaluable for their consistency
over a long time baseline of decades.  Contamination by extended
emission from the large Homunculus Nebula is always a concern in
studying Eta Car's variability.  However, high-resolution images show
that the $K$-band light is dominated by the central star to a much
higher degree than at visual wavelengths (Smith \& Gehrz 2000; Rigaut
\& Gehring 1995).  Furthermore, Smith \& Gehrz (2000) demonstrated
that near-IR variability is localized to the core region.

Figure~\ref{fig:plotJHKL} shows the $JHKL$ photometry converted to
flux units, where times of periastron passage in the binary system are
identified with vertical dashed lines, adopting a period of 2022.7
days and phase 0.0 at 2003.49 (Damineli et al.\ 2008).  One can most
clearly recognize the IR peaks associated with the two periastron
events in 1981.4 and 1986.9.  As noted previously by Whitelock et al.\
(1994, 2004), a secular increase in flux is also present in all four
filters, although to a much lesser degree in the $L$ band, presumably
due to the reduced effects of extinction at the longer wavelength and
a stronger contribution due to thermal emission from the Homunculus.
The solid curves in Figure~\ref{fig:plotJHKL} show trends that provide
a satisfactory approximation of the slow increase in each filter (see
also Whitelock et al.\ 1994, 2004).

The relative rise in flux in each band represented by these solid
curves provides an estimate of the wavelength dependence of the
thinning extinction in the foreground polar lobe of the Homunculus.
As the extinction continually decreases, the same wavelength-dependent
extinction law is used at all epochs to generate the curves in
Figure~\ref{fig:plotJHKL}.  This extinction law is shown in
Figure~\ref{fig:extinction}, where it is compared to the average
near-IR interstellar medium (ISM) extinction (dashed) from Indebetouw
et al.\ (2005).  The dust in $\eta$~Car appears to follow a different
extinction law than the normal ISM, with much higher opacity in the
$H$ and $K$ bands.  This may be a consequence of large grains that can
still scatter and absorb efficiently at 1--2 $\mu$m.\footnote{Note,
  however, that while Figure~\ref{fig:extinction} is an accurate
  representation of the wavelength dependence of the brightening,
  attributing it solely to extinction and inferring dust properties
  should be done with caution, since spatially extended thermal
  emission from the core of the Homunculus contributes to the
  $L$-band.}  There are several other lines of evidence that suggest
unusually large grains in the Homunculus, including very gray
extinction at visual wavelengths and near-blackbody temperatures
(Hillier et al.\ 2006; Smith \& Ferland 2007; Smith et al.\ 2003b).
Note that this extinction law applies to cool dust in the foreground
south-east polar lobe, and not to the hot dust proposed to be forming
during periastron passages.

The proper correction at the latest times is less certain because of
the apparent change in the rate of brightening after 1998 (e.g.,
Davidson et al.\ 1999).  We therefore interpret the 2003.5 event with
caution, and indeed, the SEDs for the most recent events may be a mix
of dust emission and wind emission.  By contrast, the IR peaks above
this trend associated with earlier events --- especially 1981.4 and
1986.9 -- are well defined and reliable, and are unambiguously due to
hot dust as discussed below.  We cannot be certain that the secular
brightening trend is due entirely to lowered extinction from the
thinning Homunculus -- it could, for example, be due in part to
long-term intrinsic changes in the central star.  In any case, we wish
to remove this trend from the data in order to focus only on the IR
excess peaks associated with the 5.5 yr cycle.  The unfilled points in
Figure~\ref{fig:plotJHKL} show the same fluxes as the filled points,
but divided by the curves noted above, resulting in a baseline flux
that appears constant over decades.

Figure~\ref{fig:plotJHKLsub} shows baseline-corrected lightcurves,
produced by dividing the light curves by the smooth secular trend to
flatten them, and then scaling their flux.  The absolute scaling is
chosen so that the most recent epochs have negligible extinction,
based on the expectation that Eta Car's line-of sight extinction is
much lower now than 30 years ago, but this choice will not affect our
results because our interest is investigating {\it relative changes}
in the SED during and after the broad peaks on time scales of $\sim$1
yr.  Therefore, we focus only on the amount of {\it excess} emission
left after subtracting off the baseline flux, shown by the unfilled
points in Figure~\ref{fig:plotJHKLsub}.  The SED of this residual
emission is the relevant quantity that we investigate in the next
section.

\begin{figure}\begin{center}
\includegraphics[width=2.6in]{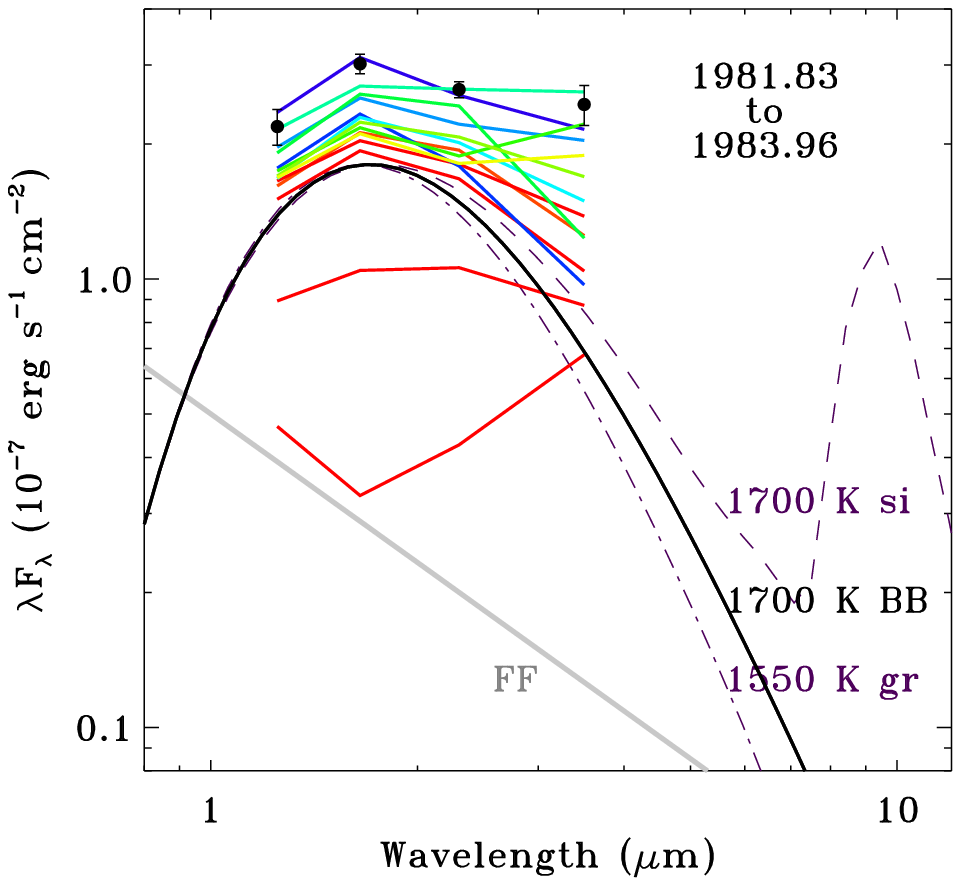}
\includegraphics[width=2.6in]{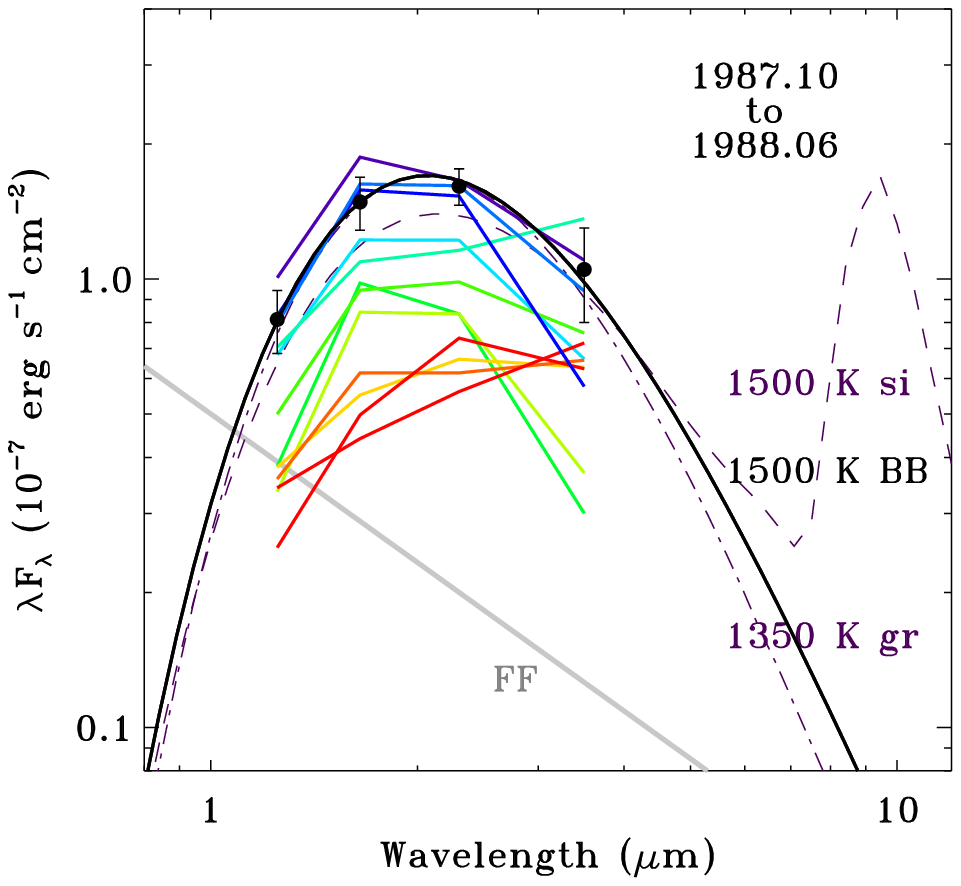}
\end{center}
\caption{Spectral energy distributions for the excess $JHKL$ emission
  associated with the IR peaks after the 1981.4 and 1986.9 periastron
  events. Plotted colors from blue/violet to red show observations for
  successive individual epochs during the time interval listed in each
  panel.  These are meant to show the range of SEDs as Eta Car faded
  after each event (see text).  The solid filled points with error
  bars are the estimates of the average flux for each event, as shown
  by the short hash marks in Figure~\ref{fig:plotJHKLsub}.  For
  comparison ({\it not} as a fit), the solid, long-dashed, and
  dot-dashed curves illustrate representative SEDs for
  single-temperature blackbodies with emissivities of $\lambda^{-1}$
  (BB), silicate (si), and graphite (gr), respectively.  The SED of
  optically thin free-free (FF) emission, with $\lambda F_{\lambda} \
  \propto \ \lambda^{-1.1}$, is shown for comparison in
  grey.}\label{fig:sed1}
\end{figure}
\begin{figure}\begin{center}
\includegraphics[width=2.6in]{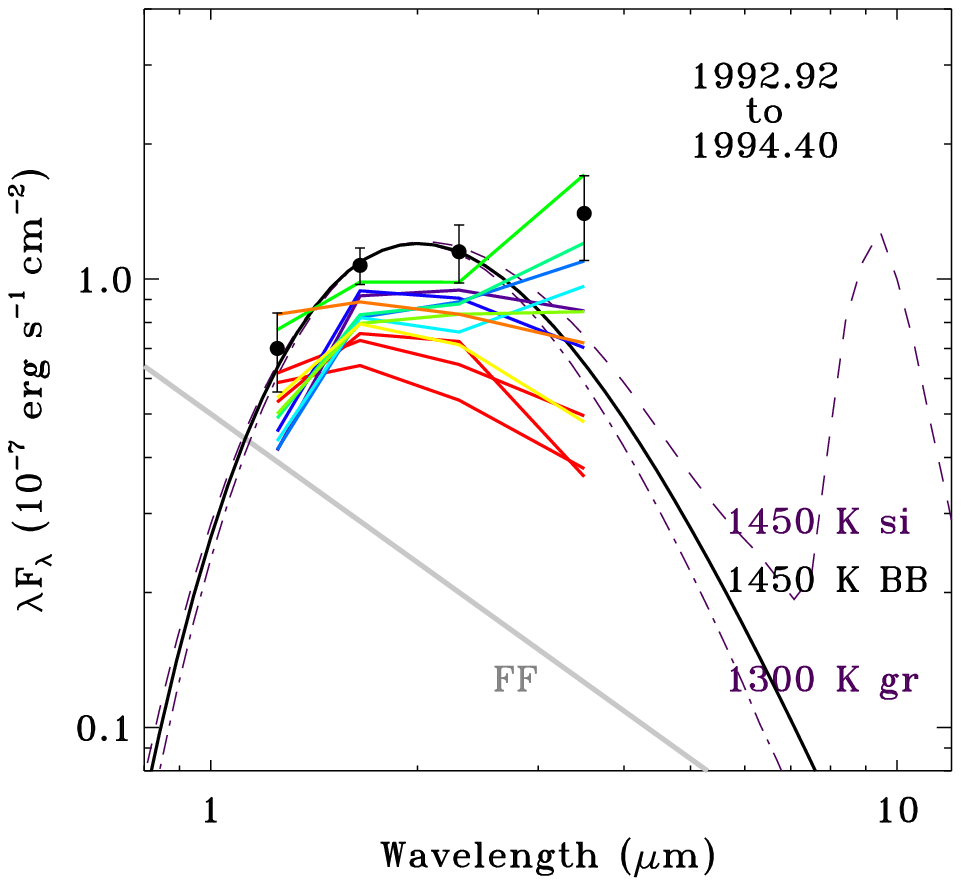}
\includegraphics[width=2.6in]{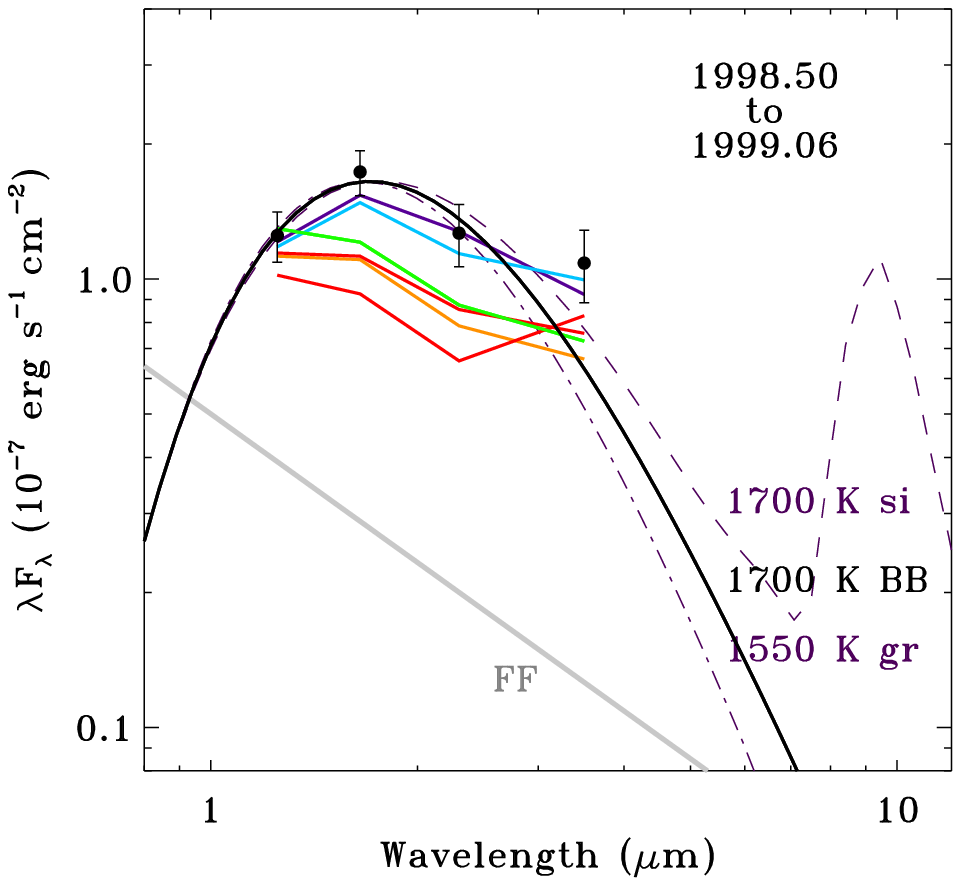}
\includegraphics[width=2.6in]{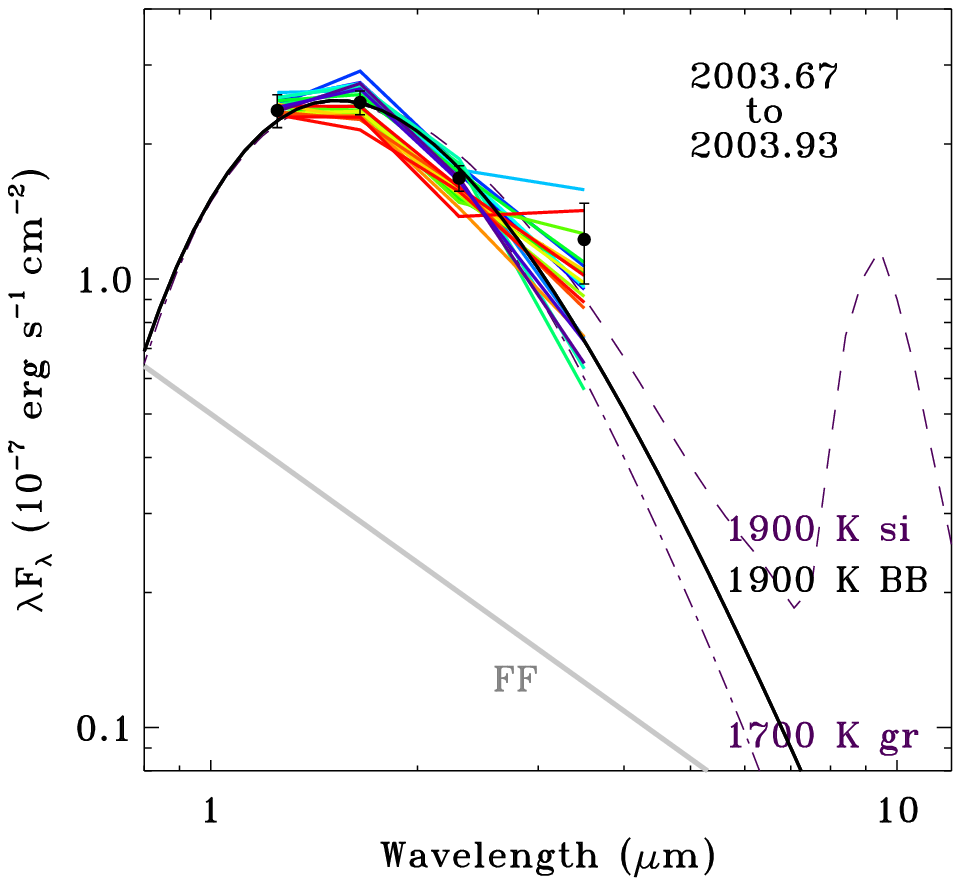}
\end{center}
\caption{Same as Figure 4, but for the 1992.4, 1998.0, and 2003.5
  events.}\label{fig:sed2}
\end{figure}

\section{SEDs OF THE NEAR-IR EXCESS}

Figures~\ref{fig:sed1} and \ref{fig:sed2} show the SEDs for residual
or {\it excess} emission associated with broad IR emission peaks after
the baseline flux has been subtracted (unfilled points in
Figure~\ref{fig:plotJHKLsub}).  Each individual panel shows a
different periastron passage event.  The solid filled points with
error bars correspond to the average flux of the broad IR excess peaks
(marked in Fig.~\ref{fig:plotJHKLsub}), while the colored SEDs
represent several individual observations for a given event, ordered
sequentially in time according to their color from blue (first) to red
(last), matching the time interval in the upper right of each plot.
For comparison, we also plot representative spectra for optically thin
free-free emission (grey), plus emission from single-temperature
grains, including a Planck function modified with emissivity
proportional to $\lambda^{-1}$ (appropriate for grains smaller than
the observed wavelength), as well as emissivities for graphite and
silicate dust with $a=$0.3~$\mu$m from Draine \& Lee (1984).  These
dust SEDs are meant for comparison and are not necessarily fits to the
data; the significance and interpretation is discussed further below.

Some mechanism gave rise to the increase in emission above Eta Car's
baseline flux during and after periastron passages, and
Figures~\ref{fig:sed1} and \ref{fig:sed2} show the spectral signature
of this excess emission for several events.  What caused it?  Most of
Eta Car's near-IR continuum flux is dominated by photospheric light
from the star plus free-free emission from the dense stellar wind,
reddened by dust to produce a smooth power-law spectrum at these
wavelengths.  However, we have already subtracted out most of this
emission and corrected for the wavelength-dependent extinction as
described above.  Emission lines from the wind and nebulosity make a
negligible contribution to the total broad-band flux (Smith 2002).

Instead, the most plausible explanation for the SED shapes in
Figures~\ref{fig:sed1} and \ref{fig:sed2} is excess emission from hot
dust.  Consider the first and most clearly-defined IR excess peak
associated with the 1981 periastron passage.  From the time of peak
emission and for $\sim$1 yr afterward, the SED (Figure~\ref{fig:sed1})
has a clear maximum in the $H$ band, and its overall shape can be
approximated quite well by modified blackbody emission at $\sim$1700~K
or somewhat cooler graphite.  Figure~\ref{fig:sed1} shows that
silicate dust at $\sim$1700~K can also match the SED shape, but this
is highly improbable since silicates have lower condensation
temperatures of 1000--1200~K.  A similar result was found for the
newly formed dust in SN~2006jc (Smith et al.\ 2008a), as discussed in
more detail below.  In Figure~\ref{fig:sed1} there is apparently some
additional flux in the $L$-band, which is likely to be due to emission
from somewhat cooler dust as well.  The type of dust which is
responsible for the near-IR excess must have a high condensation
temperature around 1700~K.  Although graphite condenses around 1600~K,
the wind of Eta Car is presumed to be C-poor, so some type of dust
other than graphite or silicate may be responsible.  Implications for
the dust composition are discussed further in \S 4.2.  The excess SED
shape cannot be explained by free-free emission or photospheric
emission, which would rise even more steeply to the blue.  After
$\sim$1 yr, the excess emission fades and the dust cools.  By $\sim$2
yr after the event (1983.96), when the peak in the light curve rejoins
the baseline flux, the hot dust signature disappears and the SED is
consistent with free-free emission and cooler dust
(Figure~\ref{sed1}).

Similarly, the excess IR emission from subsequent events is also
consistent with dust at a range of temperatures.  The peak associated
with the 1986.9 event (Figure~\ref{fig:sed1}) shows an SED that is
very well-fit by hot $\sim$1500~K dust, which then seems to cool to
$\sim$1000~K as it fades over the subsequent year.  The IR excess SEDs
from the aftermaths of the next three periastron events in 1992.4,
1998.0, and 2003.5 events are shown in Figure~\ref{fig:sed2}.  These
IR peaks were less pronounced than the previous two events
(Figure~\ref{fig:plotJHKLsub}), probably indicating a smaller mass of
newly formed dust.  The weaker dust signature may also indicate a
greater level of contamination from photospheric or free-free wind
emission, making the estimated dust temperatures for these events less
certain. Nevertheless, the red $J-H$ colors are clearly inconsistent
with free-free or photospheric emission alone, so some contribution
from dust is required.  The SED from the last event in 2003.5 ---
which by itself would imply very hot dust at 1900 K --- may be
explained by a mix of variable photospheric or free-free wind emission
combined with some hot dust, and as we noted earlier, the baseline
correction for the underlying trend is uncertain.

Zanella et al.\ (1984) also suggested that dust caused the $H$-band
excess in 1981, although they did not analyze the SED, and they
hypothesized that dust formation was due to the temporary shielding of
UV light by a shell ejection in a single-star model.  Subsequent
studies assumed that free-free emission drove the IR variability.
Chesneau et al.\ (2005) also mentioned the possibility of continuing
dust formation in the inner Homunculus, although in a different
context.

\section{DISCUSSION}

\subsection{Observed Precedents for Post-shock Dust Formation}

Although one may expect hot, X-ray emitting post-shock gas to be
hostile to the formation of grains that require temperatures below
2000~K, two different classes of object have revealed strong evidence
for the condensation of new dust grains in fast shocks.  Both classes,
discussed below, share interesting similarities with Eta Carinae.

First, some Wolf-Rayet (WR) binaries of WC type show clear evidence
for dust formation in their winds (Gehrz \& Hackwell 1974).  Perhaps
the most relevant example is the episodic dust formation in the
eccentric WC binary WR~140 (Monnier et al.\ 2002; Williams et al.\
1990; Hackwell et al.\ 1979), whose colliding winds give rise to an
X-ray light curve that is similar in several respects to that of Eta
Car (Pittard \& Dougherty 2006).  WR~137 and WR~19 are similar
examples (Williams et al.\ 2001, 2009).  There are also several cases
of WC binaries with more circular orbits that produce the so-called
``pinwheel'' nebulae.  High resolution imaging of these systems
(Tuthill et al.\ 1999; Monnier et al.\ 1999) provides strong evidence
that their IR emission comes from dust forming in post-shock gas as a
consequence of their colliding winds.  Dust may form in the cone of
post-shock gas as it flows away from the central stars and cools below
condensation temperatures.  The C-rich composition of the winds aids
the efficiency of amorphous carbon dust formation (Gehrz \& Hackwell
1974).

Second, recent studies of some supernovae (SNe) that interact with
dense circumstellar matter also provide clear evidence for the
condensation of new dust grains in the post-shock gas.  The best case
is SN~2006jc, which at only $\sim$50 days after explosion,
simultaneously showed a fading of its visual light due to extinction,
an increase in its IR flux caused by dust at $\sim$1600~K, and a
systematic blueshift of its emission lines as the far side of the SN
was blocked by the new dust.  Smith et al.\ (2008a) first demonstrated
these changes and showed that the newly formed dust must be in the
post-shock gas.  SN~2006jc is particularly relevant to Eta Car,
because at the same time that SN~2006jc formed dust and exhibited an
IR excess, it also showed a boost in its X-ray luminosity (Immler et
al.\ 2008) and enhanced He~{\sc ii} $\lambda$4686 emission (Smith et
al.\ 2008a).  He~{\sc ii} $\lambda$4686 in Eta Car is also associated
with the periastron passages when dust forms (Steiner \& Damineli
2004; Martin et al.\ 2006), as is enhanced X-ray emission.  As noted
by Smith et al.\ (2008a), these similarities implicate common physical
conditions in the post-shock cooling zones of Eta Car and SN~2006jc.

While both WC binaries and SN~2006jc have C-rich material that may aid
the formation of graphite grains, this is not necessarily a
prerequisit for dust formation in post-shock gas.  There is now a
growing number of additional cases for post-shock dust formation in
Type IIn supernovae that are {\it not} expected to have C-rich ejecta
(Smith et al.\ 2009, 2008b; Fox et al.\ 2009; Pozzo et al.\ 2004).
Details of how dust forms in the hot post-shock gas is not yet fully
understood, but these observed examples prove that Nature can overcome
the difficulties involved.  Strong radiative cooling in dense
post-shock gas is probably instrumental (Usov 1991), as are density
inhomogeneities.

\subsection{Post-Shock Grain Condensation in the Colliding Winds of
  Eta Car}

Given the episodic presence of the hot dust in Eta Car, the most
promising place for dust to form is in the compressed post-shock
primary wind, akin to the episodic dust formation in WR~140 and other
colliding-wind binary systems discussed above.  Recent hydrodynamic
simulations of the colliding winds in Eta Car (Parkin et al.\ 2009;
Okazaki et al.\ 2008) show dense gas clumps arising from instabilities
at the colliding-wind shock, with $n_H > 10^{12}$ cm$^{-3}$ at radii
of 2--3 AU around times of periastron.  These clumps will cool and
become somewhat less dense as they flow down the shock cone,
eventually reaching a radius where the radiative equilibrium
temperature drops low enough for grains to condense.  For sublimation
temperatures around $T\simeq$1700~K, this occurs at $R \ \simeq \
60$~AU in Eta Car, which is only 2--3 times the orbital separation at
apastron, and for which the flow timescale is only a few months at
$\sim$500 km s$^{-1}$.  Densities of clumps at these radii may drop to
$\sim$10$^9$ cm$^{-3}$, comparable to minimum densities for grain
nucleation (e.g., Smith et al.\ 2008a and references therein).  Grains
may form at even smaller radii if high optical depths shield clumps
from the primary star's radiation.

Once the dust grains have formed, they will cool quickly as they
expand down the shock cone.  Flowing outward at 500 km s$^{-1}$ from
an initial formation radius at 60 AU, for example, the dust
equilibrium temperature will fall to $\sim$1400 K after 100 days and
will drop to $\sim$1000 K after 1 yr.  This roughly matches the dust
cooling indicated by the sequence of SEDs observed in the year
following the 1987 periaston event (Fig.~\ref{fig:sed1}).  The
observed dust temperature does not necessarily need to drop however,
because new dust may continue to form at the same temperature in the
shock cone, replacing the dust that has cooled.  Some other events
have SEDs showing relatively constant dust temperatures.

Long after the periaston event when the new dust has cooled, it should
reside primarily in the complex expanding shells and loop structures
created during the 5.5 yr orbit, like those seen in simulations of Eta
Car's colliding winds (Parkin et al.\ 2009; Okazaki et al.\ 2008).  It
would be quite interesting to make a detailed comparison between the
geometry of these simulations and dust structures observed in the core
of the Homunculus.  High-resolution data show an intriguing asymmetric
distribution of material around the star (e.g., Smith et al.\ 2004;
Chesneau et al.\ 2005; Gull et al.\ 2009), in which high grain
temperatures of 400--500~K are observed (Smith et al.\ 1998, 2003b;
Hackwell et al.\ 1986).

Some evidence suggests that Eta Car's wind mass-loss rate changes from
one cycle to the next (Damineli et al.\ 1998, 2008; Davidson et al.\
2005), so the relevant post-shock density may be different in each
cycle.  Depending on how efficiently the gas cools, when dust starts
to form, and how long the dust formation episode lasts, significantly
different IR excess emission may be observed in each successive
periastron event, consistent with the somewhat irregular character of
the $JHKL$ light curves.  With peak $\lambda F_{\lambda}$ values of
(1--3)$\times$10$^{-7}$ erg s$^{-1}$ cm$^{-2}$, the luminosity of the
hottest dust that dominates the excess $JHK$ fluxes in
Figures~\ref{fig:sed1} and \ref{fig:sed2} is roughly
(4--12)$\times$10$^4$ $L_{\odot}$ (i.e. a small fraction of the star's
radiative energy budget, but far exceeding the X-ray luminosity).  At
temperatures of roughly 1400--1700~K, the dust mass needed to account
for the IR luminosity is at least (1.4--2.8)$\times$10$^{-8}$
$M_{\odot}$ (adopting the same simplifying assumptions and methods as
Smith \& Gehrz 2005).  This is a lower bound to the mass formed in a
periastron event because cooler dust is also present, as is evident
from the SED shapes --- the hottest dust cools quickly as it flows
downstream in the shock cone and may be replaced by additional hot
dust (see Smith et al.\ 2008a).  This is less than 0.1\% of the total
gas mass entering the shock interaction cone during the year around
periastron.

Both the dust mass and the fraction of wind mass converted to dust in
one of Eta Car's periastron events are quite similar to values
estimated for WR~140, which are 2.8$\times$10$^{-8}$ $M_{\odot}$ and
0.2\%, respectively (Williams et al.\ 1990).  Broad peaks in the
multi-filter IR light curves of WR~140 (Williams et al.\ 1990) are
also similar to those of Eta Car.

In dusty WC binaries and SN~2006jc, gas entering the shock is expected
to be C enriched, favoring the formation of amorphous carbon grains.
In Figures~\ref{fig:sed1} and \ref{fig:sed2}, we have compared the
observed SEDs to graphite and silicates, as well as a Planck function
with a more generic $\lambda^{-1}$ emissivity that may be
representative of some other grain species.  There are problems with
attributing the new dust to either graphite or silicate grains.
First, the high temperatures implied by the SED shapes are
inconsistent with silicates and Fe-dominated grains, which seem to
condense at lower temperatures of 1000--1200~K in novae (e.g., Gehrz
1988).  While some silicates may form and may be partly responsible
for the cooler dust excess seen in the $L$-band, silicates cannot
dominate the hottest $\sim$1700~K dust seen at $JHK$.  The high
temperatures are in principle consistent with graphite, which can
condense at temperatures around 1600~K, but that would be surprising
given the N-rich but severely O/C-poor abundances observed in Eta Car
(Davidson et al.\ 1986).  The dust mass formed in a periastron event
is small compared to the available gas mass, so while C should be
underabundant, the formation of some small amount of amorphous carbon
is not necessarily precluded.

A more straightforward solution, however, may be to attribute most of
the newly formed hot dust to another species, such as corundum
(Al$_2$O$_3$), which condenses at high temperatures around 1700~K.
Indeed, Chesneau et al.\ (2005) argued that some contribution from
corundum was needed to fit the very wide mid-IR ``silicate'' feature
in Eta Car, and they found some spatial variation in the relative
corundum abundance in the core of the Homunculus.  In particular,
Chesneau et al.\ (2005) found that a clump of dust located 0$\farcs$4
southeast of the central star had a higher relative abundance of
corundum than dust features in other directions, which they attributed
to dust formation in a polar wind.  Smith (2006) showed that this SE
clump was redshifted and therefore located on the far side of the star
in the equatorial plane.  The fact that this equatorial clump is
enhanced in corundum is quite interesting in the current context,
because the orbit orientation favored by Okazaki et al.\ (2008) places
this clump in the direction of the secondary star at periastron
passages.  In other words, if corundum dust forms preferentially
around times of periaston in connection with the IR peaks in the
photometry, and it forms in the colliding wind shock cone, then {\it
  this is the preferential direction in which we would expect the new
  dust to flow}.  In WR~140, for example, puffs of episodic dust
formed in the shock cone at periaston expand away from the star in
that direction in high spatial resolution images (Monnier et al.\
2002).  This hints that further study of the spatial structure and
composition of the dust in the core of Eta Car may prove to be
extremely interesting.

\subsection{Summary and Further Implications}

Following the precendent set by dusty WC binaries that form dust in
their colliding-wind shocks and by some SNe that form dust in
post-shock gas, this paper proposes that the broad peaks in the $JHKL$
light curve of Eta Car are due to the post-shock condensation of new
dust in the colliding winds.  The dust forms around the time of
periastron passage in the eccentric 5.5 yr binary orbit, probably
within dense clumps that have expanded far enough from the star that
the radiative equlibrium temperature drops below $\sim$1700~K, and the
dust then cools as it flows down the shock cone.  The clearest
episodes were associated with with the 1981.4 and 1986.9 periastron
events, although later events also show evidence for dust.  Episodic
formation of dust in the colliding winds of the Eta Car binary system
has several additional implications for this complex system:

1) Dust formed recently in this way is near the central star (within
60--500 AU), compared to the dust in the much larger Homunculus
Nebula.  While it is a relatively small mass of dust, its proximity
may cause substantial extinction.  This dust may preferentially block
our view of the star without obscuring features in the surrounding
nebulosity and extended wind, perhaps explaining Eta Car's mysterious
``coronograph'' described by Hillier et al.\ (2006).

2) Absorption and scattering by this localized dust may be highly time
variable as the dust expands and thins in the time interval between
periastron passages.  This dust may block starlight, or may scatter
starlight toward us, with different contributions at different times.
It may therefore contribute to irregular variability in optical
photometry of Eta Carinae.  Sterken et al.\ (1998) noted a drastic
reddening of visual colors after several events, which they attributed
to (usually slower) S~Doradus variations, but perhaps the sudden
reddening was from new dust close to the star instead.

3) Combined effects of local scattering and extinction may also
influence spectra, even on very small angular scales of
0$\farcs$1--0$\farcs$2.  The asymmetric expanding dust formed in the
colliding-wind shock cone could serve as a moving mirror, reflecting
some specific lines of sight more than others.  This may be a factor
in perplexing subtleties of spatially-dependent emission line profiles
along our line of sight (e.g., Gull et al.\ 2009; Davidson et al.\
2005), and direction-dependent profiles reflected by the Homunculus
(e.g., Smith et al.\ 2003a).  Even line profiles in our direct view of
the central star may be affected by time-dependent dust scattering and
extinction.

4) Episodic dust formation at small radii may regulate the escape of
UV photons that excite emission lines in nearby gas condensations,
contributing to the observed nebular spectral variability during the
5.5 yr cycle, and mimicing a shell ejection as described by Zanella et
al.\ (1984).

5) Dust production in successive events every 5.5 yr is not perfectly
repeatable, signaling differences in the post-shock densities that may
result from longer-term changes in the mass-loss rate of the primary
star.  This, in turn, may alter the effects discussed above from one
cycle to the next.

6) Episodic post-shock dust formation in the colliding winds of Eta
Carinae may favor different refractory materials than those which
condensed in the polar lobes.  The more massive Homunculus may have
had much higher densities and lower temperatures following the 19th
century Great Eruption, fostering efficient formation of silicates, Fe
grains, or other species that require lower temperatures.  The
transitory nature of the dust formation in colliding winds, however,
may only provide sufficiently high densities for a brief time when the
temperatures are high, permitting only corundum, amorphous carbon, or
other species that condense at high temperatures.  The limited
duration of dust formation could then preclude the formation of lower
temperature condensates.  This might lead to spatial variations in
dust composition and grain size around Eta Carinae, of which there are
some observations signatures (e.g., Robinson et al.\ 1987; Smith et
al.\ 1998; Chesneau et al.\ 2005).

\smallskip

It would obviously be interesting to spatially resolve the expanding
dust that formed in recent eposides in order to constrain its
geometry, as has been done for WR systems mentioned earlier, and to
obtain mid-IR spectra of this newly formed dust as it expands and
cools.  This may prove a difficult challenge amid the bright and
spatially complex Homunculus, however.  Dust formation now may be less
efficient than it was during 1981--1987, due to changes in the primary
star's mass-loss.

\smallskip\smallskip\smallskip\smallskip
\noindent {\bf ACKNOWLEDGMENTS}
\smallskip
\scriptsize

I am indebted to P.A.\ Whitelock for providing electronic tables of
the $JHKL$ photometry obtained at the SAAO from 1972 to 2004, and I
thank the referee, P.\ Williams, for helpful comments that improved
the paper.  I was partially supported by NASA through grants GO-10241
and GO-10475 from the Space Telescope Science Institute, which is
operated by the Association of Universities for Research in Astronomy,
Inc., under NASA contract NAS5-26555.


\end{document}